\documentclass[aps,prl,onecolumn,superscriptaddress,preprint,preprintnumbers]{revtex4}

\usepackage{graphicx}
\usepackage{dcolumn}
\usepackage{bm}
\usepackage{amsmath}
\usepackage{color,soul}
\usepackage{upgreek}
\usepackage{xcolor}
\usepackage{url}

\newlength{\figurewidth}
\begin{document}


\title{Ultrafast and energy-efficient quenching of spin order:\\ Antiferromagnetism beats ferromagnetism}

\author{Nele Thielemann-K{\"u}hn}
\email{nele.thielemann@helmholtz-berlin.de}
\altaffiliation[Present address: ]{Fachbereich Physik, Freie Universit{\"a}t Berlin, Arnimallee 14, 14195 Berlin, Germany }%
\affiliation{Institut f{\"u}r Methoden und Instrumentierung der Forschung mit Synchrotronstrahlung, 
	Helmholtz-Zentrum Berlin f{\"u}r Materialien und Energie GmbH, Albert-Einstein-Stra{\ss}e 15, 12489 Berlin, Germany}%
\affiliation{Institut f{\"u}r Physik und Astronomie, Universit{\"a}t Potsdam, Karl-Liebknecht-Stra{\ss}e 24/25, 14476 Potsdam, Germany}%

\author{Daniel Schick}
\affiliation{Institut f{\"u}r Methoden und Instrumentierung der Forschung mit Synchrotronstrahlung, 
	Helmholtz-Zentrum Berlin f{\"u}r Materialien und Energie GmbH, Albert-Einstein-Stra{\ss}e 15, 12489 Berlin, Germany}%

\author{Niko Pontius}
\affiliation{Institut f{\"u}r Methoden und Instrumentierung der Forschung mit Synchrotronstrahlung, 
	Helmholtz-Zentrum Berlin f{\"u}r Materialien und Energie GmbH, Albert-Einstein-Stra{\ss}e 15, 12489 Berlin, Germany}%

\author{Christoph Trabant}
\affiliation{Institut f{\"u}r Methoden und Instrumentierung der Forschung mit Synchrotronstrahlung, 
	Helmholtz-Zentrum Berlin f{\"u}r Materialien und Energie GmbH, Albert-Einstein-Stra{\ss}e 15, 12489 Berlin, Germany}%
\affiliation{Institut f{\"u}r Physik und Astronomie, Universit{\"a}t Potsdam, Karl-Liebknecht-Stra{\ss}e 24/25, 14476 Potsdam, Germany}%
\affiliation{II. Physikalisches Institut, Universit{\"a}t zu K{\"o}ln, Z{\"u}lpicher Stra{\ss}e 77, 50937 K{\"o}ln, Germany}%

\author{Rolf Mitzner}
\affiliation{Institut f{\"u}r Methoden und Instrumentierung der Forschung mit Synchrotronstrahlung, 
	Helmholtz-Zentrum Berlin f{\"u}r Materialien und Energie GmbH, Albert-Einstein-Stra{\ss}e 15, 12489 Berlin, Germany}%

\author{Karsten Holldack}
\affiliation{Institut f{\"u}r Methoden und Instrumentierung der Forschung mit Synchrotronstrahlung, 
	Helmholtz-Zentrum Berlin f{\"u}r Materialien und Energie GmbH, Albert-Einstein-Stra{\ss}e 15, 12489 Berlin, Germany}%

\author{Hartmut Zabel}
\affiliation{Institut f{\"u}r Physik, Johannes-Gutenberg-Universit{\"a}t Mainz, Staudingerweg 7, 55128 Mainz, Germany}%

\author{Alexander F{\"o}hlisch}
\affiliation{Institut f{\"u}r Methoden und Instrumentierung der Forschung mit Synchrotronstrahlung, 
	Helmholtz-Zentrum Berlin f{\"u}r Materialien und Energie GmbH, Albert-Einstein-Stra{\ss}e 15, 12489 Berlin, Germany}%
\affiliation{Institut f{\"u}r Physik und Astronomie, Universit{\"a}t Potsdam, Karl-Liebknecht-Stra{\ss}e 24/25, 14476 Potsdam, Germany}%

\author{Christian Sch{\"u}{\ss}ler-Langeheine}
\affiliation{Institut f{\"u}r Methoden und Instrumentierung der Forschung mit Synchrotronstrahlung, 
	Helmholtz-Zentrum Berlin f{\"u}r Materialien und Energie GmbH, Albert-Einstein-Stra{\ss}e 15, 12489 Berlin, Germany}%

\date{\today}

\begin{abstract}
By comparing femtosecond laser pulse induced ferro- and antiferromagnetic dynamics in one and the same material - metallic dysprosium - we show both to behave fundamentally different. 
Antiferromagnetic order is considerably faster and much more efficiently reduced by optical excitation than its ferromagnetic counterpart. 
We assign the fast and extremely efficient process in the antiferromagnet to an interatomic transfer of angular momentum within the spin system.
Our findings imply that this angular momentum transfer channel is effective in other magnetic metals with non-parallel spin alignment. They also point out a possible route towards energy-efficient spin manipulation for magnetic devices.
\end{abstract}


\maketitle

Striving for novel concepts for faster and more energy-efficient data processing and storage, a wealth of experimental and theoretical studies in the field of ultrafast magnetic dynamics has been carried out \cite{Kirilyuk2010,Steiauf2009,Koopmans2010,Schellekens2013,Mueller2011,Battiato2010,Malinowski2008,Rudolf2012,Eschenlohr2013,Essert2011,Roth2012,Ostler2012,Bergeard2014,Wietstruk2011}. 
This entailed the understanding that a speed limit for spin manipulation is governed by the achievable angular momentum transfer. 
For any change of magnetic order fundamental conservation laws require transfer of angular momentum associated with the atomic magnetic moments \cite{Kirilyuk2010,Einstein1915}. 
This is particularly relevant when magnetic order is to be affected on ultra-short time scales, e.g., by femtosecond laser-pulse excitation. 
Here the angular momentum transfer effectively limits the speed of magnetic dynamics. 
Various transfer channels have been identified including local scattering processes \cite{Steiauf2009,Koopmans2010,Schellekens2013,Mueller2011} as well as spin transport \cite{Battiato2010,Malinowski2008,Rudolf2012,Eschenlohr2013}, and their relative importance for ultrafast magnetic dynamics is subject of intense debate \cite{Kirilyuk2010,Koopmans2010,Essert2011,Roth2012,Ostler2012}.
Changing ferromagnetic (FM) order via local processes requires angular momentum transfer out of the spin system into an external reservoir like the lattice. 
In contrast, the change of antiferromagnetic (AFM) order with vanishing net magnetization, could be achieved by redistribution of angular momentum within the spin system itself; transfer of angular momentum into other degrees of freedom is not required.
One would therefore expect any change of AFM order to occur faster than modifications of FM order.

So far, AFM dynamics was mostly studied experimentally in transition-metal oxides and it was found to proceed over a wide range of time scales including ultrafast dynamics within 230\,fs \cite{Fiebig2008}, but also much slower dynamics on picosecond time scales \cite{Tobey2012}. 
In ferrimagnetic metallic alloys of 3$d$ and 4$f$ metals, ultrafast angular momentum transfer between antiferromagnetically exchange-coupled sublattices was observed \cite{Bergeard2014,Radu2011}. 
These results, however, are not straightforwardly comparable to the wealth of work about FM metals: for 3$d$-4$f$ alloy dynamics static inhomogeneity has been shown to play a crucial role \cite{Graves2013}; and in oxides the exchange coupling mechanisms are different to those in metals. 
This renders quantitative comparison with the thoroughly studied elemental ferromagnets ambiguous. 
Already within one material class any magnetic dynamics - FM as well as AFM - is expected to depend on the size of the magnetic moment \cite{Radu2015} and on material properties like the spin-orbit, spin-lattice and electron-lattice interaction \cite{Kirilyuk2010}. To avoid such complications, we compare FM and AFM dynamics in the most direct way in one and the same material: metallic dysprosium (Dy).

\begin{figure*}[!htb]
	\centering
	\includegraphics[width=0.9\textwidth]{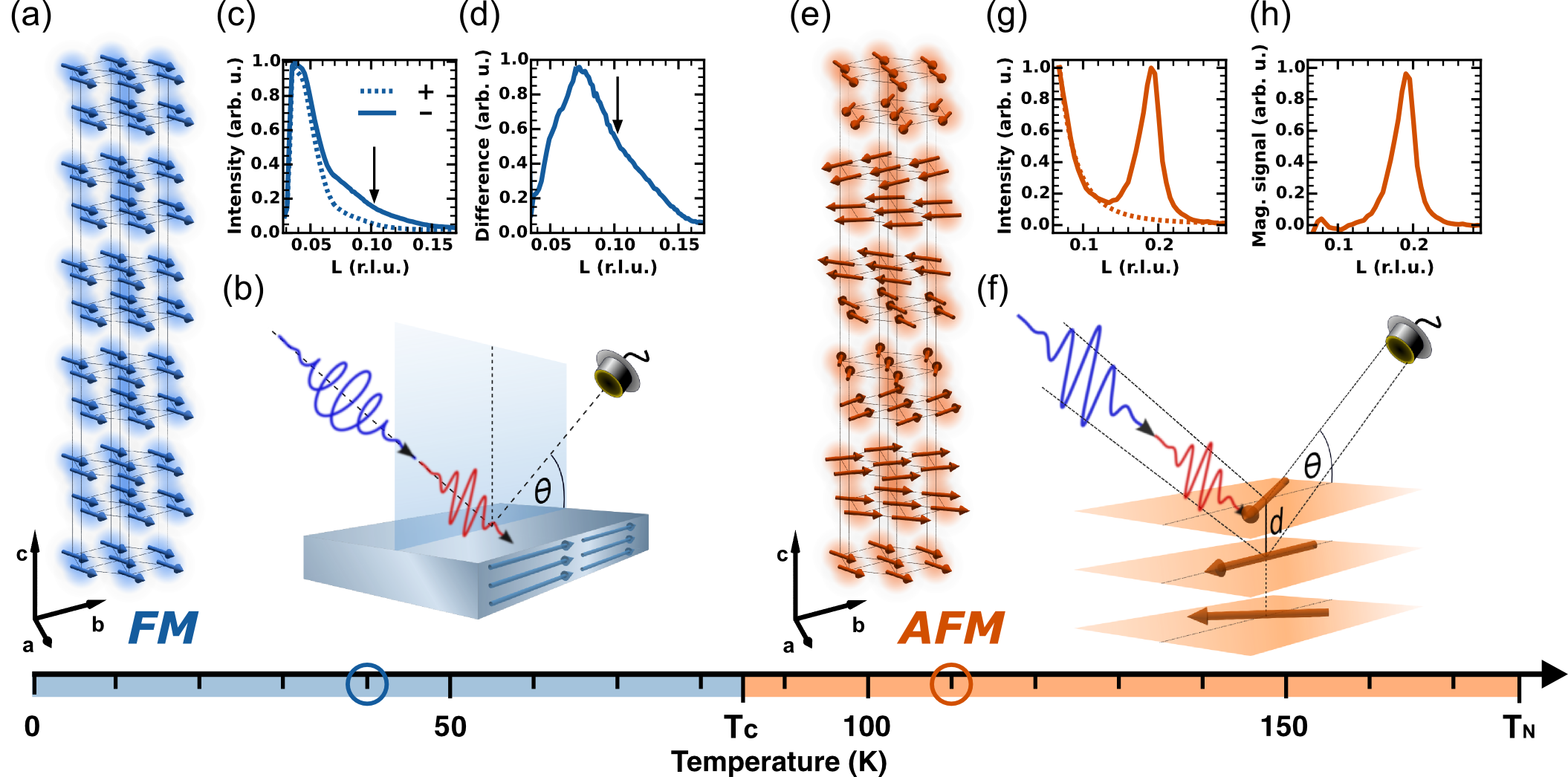}
	\caption{(a) FM structure of the 4$f$ spins. 
	    (b) Geometry for probing ferromagnetism with circularly polarized soft x-rays tuned to the Dy $M_5$-resonance; a near-infrared laser pulse is shown as red wave packet. 
		(c) Specular reflectivity vs. momentum transfer $L$, for opposite photon-helicity projections on the sample magnetization (solid and dashed lines). 
		The difference (d) is the FM contrast and the temporal response is probed at the momentum transfer value marked by the arrow.
	    (e) AFM spin structure.
	    (f) Geometry for probing AFM order with linearly polarized soft x-rays. 
		(g) Magnetic Bragg peak due to the magnetic helix period length (about 5 times the crystalline unit cell or 10 atomic layers) located on a weak reflectivity background (dashed line) that has been subtracted in (h).		
	}
	\label{fig:overview}
\end{figure*}

Dy is FM at low temperatures and has a helical AFM phase between 85\,K and 178\,K \cite{Behrendt1958}, see Fig.\,\ref{fig:overview}. 
The strongly localized 4$f$ magnetic moments (10\,$\upmu$B per atom) are magnetically coupled by indirect (RKKY) exchange through intra-atomic spin polarization of mostly 5$d$ states in the (5$d$6$s$) conduction band \cite{jensen1991rare,Ahuja1994}.
AFM and paramagnetic Dy has $hcp$ symmetry; the FM phase shows an orthorhombic distortion \cite{darnell1963}.
In the FM phase all 4$f$ spins are parallel aligned within the basal $ab$-planes, see Fig.\,\ref{fig:overview}\,(a).
In the AFM phase the 4$f$ spins within each $ab$-plane remain ferromagnetically aligned but form an helical structure along the crystallographic $c$-axis, see Fig.\,\ref{fig:overview}\,(e).

FM and AFM 4$f$ order can straightforwardly be probed with soft x-rays tuned to the 3$d$~$\rightarrow$~4$f$ electronic excitation ($M_5$-edge at around 1292\,eV photon energy).
For FM order we used magnetic circular dichroism (MCD) in reflection geometry \cite{Mertins2002, Macke2014}, i.e., the effect that a FM sample reflects elliptically polarized x-rays differently depending on the photon helicity projection onto the sample magnetization.
For probing FM dynamics the sample was held at 40\,K. 
The specular reflected intensity at the maximum of the Dy $M_5$-absorption edge was recorded for opposite sign of a magnetic field of 80\,mT oriented in the scattering plane and parallel to the sample surface. 
Incidence and detection angles were set to 5~degrees with respect to the sample surface. 
In order to determine the FM order parameter the difference in reflected intensities for opposite direction of the magnetic field was taken \cite{Mertins2002}.

AFM order was studied by resonant magnetic x-ray diffraction: The helical magnetic order leads to a superstructure Bragg peak \cite{Ott2010} at (0\,0\,$\tau$) with $\tau \approx 0.19$ in reciprocal lattice units (r.l.u.).
Data in the AFM phase were recorded using linearly polarized x-rays with the sample held at 110\,K. 
The magnetic diffraction peak at (0\,0\,$\tau$) occurs in specular geometry with an incidence angle of about 9.5~degrees with respect to the sample surface. 
In order to determine the AFM order parameter the square root of scattering signal was calculated \cite{Ott2006Ho}.

Since resonant magnetic x-ray diffraction and magnetic circular dichroism are based on exactly the same contrast mechanism \cite{Lovesey1996}, a combination of both techniques allows for determining the FM and AFM order parameters in a directly comparable way.
For an overview of experimental geometries and data acquisition see Fig.\,\ref{fig:overview}.
As sample we chose a 120\,nm thin metallic Dy film grown by molecular beam epitaxy with (0\,0\,1) surface orientation. The film was sandwiched between Yttrium (Y) layers to minimize strain; Niobium (Nb) served as buffer layer and oxidation protection; sapphire was the substrate \cite{Leiner2004}.
The stacking in the film was Nb\,(2.5\,nm)\,/\,Y\,(3\,nm)\,/\,Dy\,(120\,nm)\,/\,Y\,(70\,nm)\,/\,Nb\,(50\,nm)\,/\,$a$-plane sapphire.

All experiments were carried out at the FemtoSpeX slicing facility at the electron storage ring BESSY\,II of the Helmholtz-Zentrum Berlin \cite{Holldack2014}. 
Magnetic dynamics was induced by 800\,nm-near-infrared-laser pulses of 50\,fs duration. 
The magnetic signal was probed with 100\,fs-x-ray pulses, hitting the sample with 6\,kHz repetition rate, while the pump laser was operated at 3\,kHz such that alternating signals with and without pump-laser excitation were detected. 
The latter were used for normalization.
The overall temporal resolution was about 120\,fs.
For our geometry the penetration depth for pump photons is about 21~nm \cite{Adachi2012}; the x-ray probing depth is 7~nm (12~nm) for the FM (AFM) case (see supplemental material \cite{Supps2017}). The probed volume in our experiment was thus fully excited by the laser.
Detailed information on the experimental setup, data acquisition, data analysis, as well as the complete set of evaluated data can be found in the supplemental material \cite{Supps2017}.


\begin{figure}
	\centering
	\includegraphics[width=0.8\columnwidth]{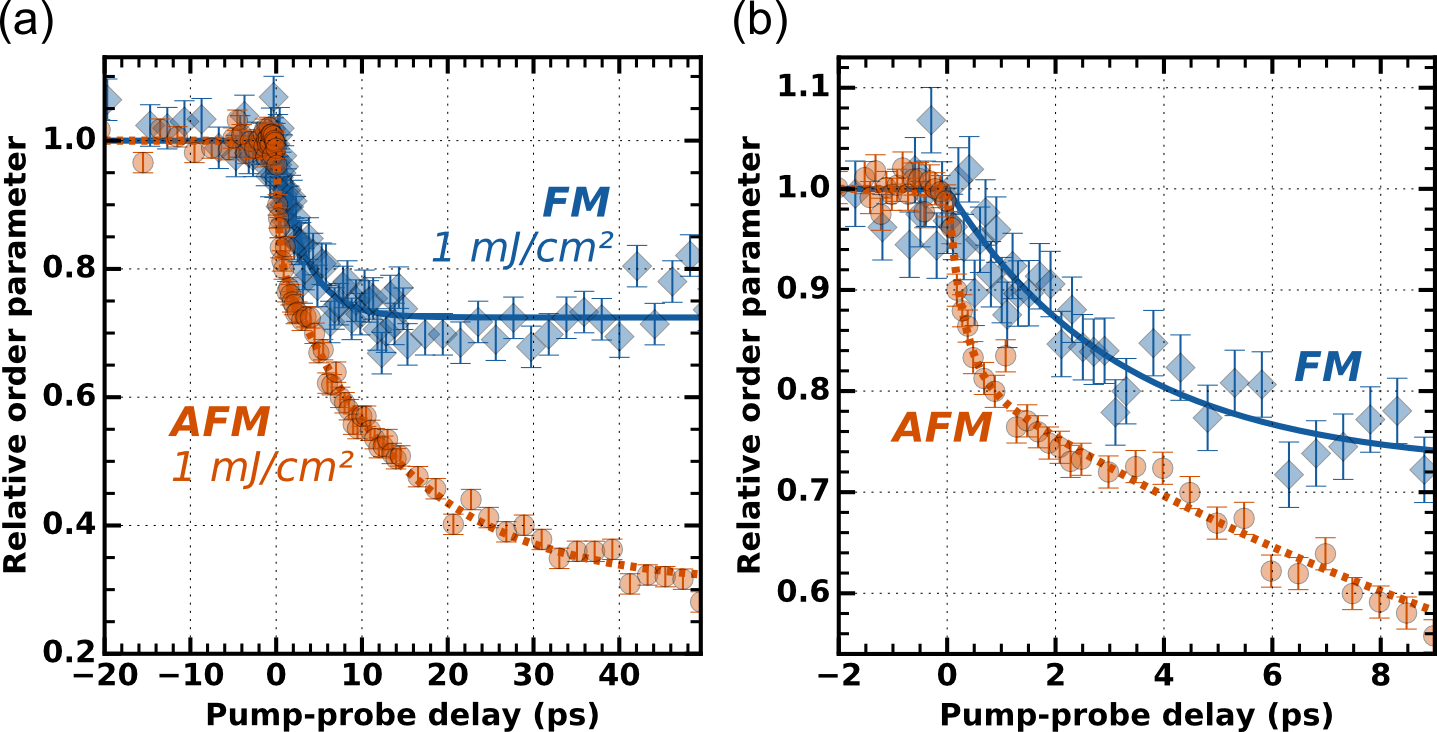}
	\caption{
		(a) Pump-probe delay scans in the FM (blue) and AFM (orange) phase for an absorbed laser fluence of 1\,mJ/cm$^2$. 
		The symbols denote the normalized magnetic order parameter;
		the lines denote exponential fits to the data. 
		(b) First 9\,ps of the delay traces on an enlarged scale. 
	}
	\label{fig:delay}
\end{figure}

Typical dynamical data are presented in Fig.\,\ref{fig:delay}\,(a).
The two transients demonstrate the clearly different response of the two order parameters. 
After an equally strong laser excitation (absorbed fluence), both magnetic order parameters are reduced, but the quenching of the AFM order is considerably and consistently stronger for all delays. 
Moreover, the shapes of the two transients are significantly different. 
Zooming into the first 9\,ps [Fig.\,\ref{fig:delay}\,(b)] reveals the initial AFM order parameter loss to occur much faster than its FM counterpart. 
For AFM dynamics in Fig.\,\ref{fig:delay} we find an initial fast reduction with an exponential time constant of ($290\,\pm\,40$)\,fs followed by a slower one of ($14\,\pm\,1$)\,ps. 
In contrast, the FM dynamics occurs with a single time constant of ($3.2\,\pm\,0.3$)\,ps.
The lines in Fig.\,\ref{fig:delay}\,(a,b) show results of least square fits to double or single exponential decay models (see supplemental materials \cite{Supps2017}). 

\begin{figure}
	\centering
	\includegraphics[width=0.8\columnwidth]{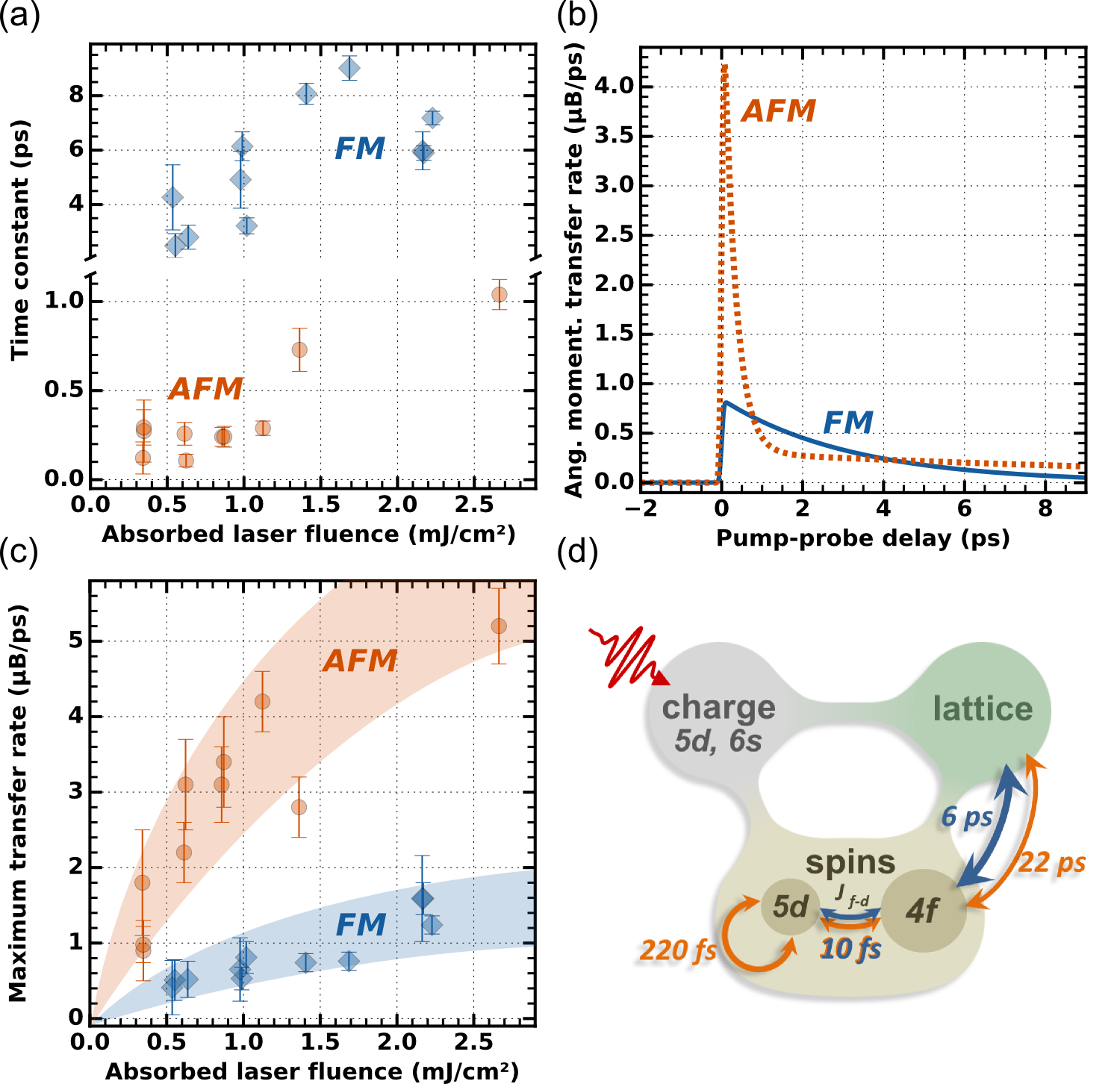}
	\caption{ 
		(a) Short time constants for the AFM and single time constants for the FM dynamics, determined from the delay traces for different absorbed laser fluences (note broken $y$-axis). 
		(b) The momentary rate of average atomic angular momentum transfer derived from the exponential fits in Fig.\,\ref{fig:delay}\,(a). 
		(c) Maximum momentary angular momentum transfer rate deduced from the delay traces for different absorbed laser fluences. 
		The shaded areas are guides to the eye. 
		(d) Channels of angular momentum transfer active in the AFM (orange arrows) and FM (blue arrows) phase of Dy (see text). 
	}
	\label{fig:results}
\end{figure}

A fluence dependent investigation, see Fig.\,\ref{fig:results}\,(a), shows the initial AFM decay time constant to vary very little for low absorbed fluences up to 1.2\,mJ/cm$^2$ with an average value of ($220\,\pm\,70$)\,fs. 
For higher fluences, the decay becomes slower, reading 1040\,fs for the highest fluence considered in this work. 
Remarkably, all initial AFM dynamics are significantly faster than the single time constants in the FM phase; the latter ones on average amount to ($6\,\pm\,2$)\,ps [Fig.\,\ref{fig:results}\,(a)]. 
The difference between FM and AFM dynamics becomes even more pronounced comparing the momentary rate of atomic angular momentum transfer [Fig.\,\ref{fig:results}\,(b)]. 
We define the (momentary) angular momentum transfer rate as the change of the magnetic order parameter per time.
The maximum transfer rate in the AFM phase is more than five times higher than in the FM phase. 
This trend is true for a wide range of laser-excitation fluences. 
In Fig.\,\ref{fig:results}\,(c) we present the maximum measured angular momentum transfer rate vs. the absorbed laser fluence. 
The maximum AFM transfer rates are always higher by a factor of four to five. 
Ultrafast reduction of spin order in the antiferromagnet is hence more energy-efficient than in the ferromagnet.

We note that in between the base temperatures of the experiments (40\,K and 110\,K, respectively) the static order parameter changes by less than 15\,\%. 
This small change can not cause such different magnetic dynamics. The higher energy efficiency along with the faster spin dynamics for the AFM phase is a true consequence of the different spin structures. 

In principle, the energy deposited by the laser above a fluence of about 1\,mJ/cm$^2$ for the 
	FM and about 1.8\,mJ/cm$^2$ for the AFM case \cite{griffel1956} would be sufficient to heat the 
	sample across the nearest phase transition. We find, however, no indications for such an effect within the time window of 
	our experiment. The clearest indication for the absence of an equilibrium phase transition is the residual FM and 
	AFM order parameter we find even after 50 ps delay and for fairly high fluences (see supplemental materials \cite{Supps2017}); 
	an equilibrium phase transition would lead to a complete loss of the respective order parameter. In fact, a long lasting non-equilibrium between Dy spin system and lattice after photoexcitation was also observed in a recent structural dynamics study \cite{Reppert2016}.

In the following we discuss the angular momentum transfer channels responsible for the observed behavior. 
We assign the difference between FM and AFM dynamics for short delay times to an angular momentum transfer channel only effective in magnetic systems like antiferromagnets, i.e., where spin orientations are not parallel. 
This transfer channel essentially bases on interatomic spin hopping. 

Since a direct excitation of 4$f$ electrons (3.8\,eV binding energy) or a transition of 5$d$6$s$ electrons into unoccupied 4$f$ states (2\,eV above Fermi level) is not possible in both magnetic phases with the pump photon-energy of 1.5\,eV \cite{Lebegue2006}, the pump-laser pulse essentially excites delocalized 5$d$6$s$ electrons.
In the AFM phase these excited electrons with their spins initially aligned parallel to the local 4$f$ spins hop to adjacent sites with non-parallel 4$f$ spins. 
This brings about a disordering of the 5$d$-spin subsystem.
Subsequently this disorder is imposed onto the 4$f$ subsystem via the strong 4$f$-5$d$ coupling \cite{Wietstruk2011}. 
Note that such interatomic spin transfer also occurs in the FM phase but - owing to the allover parallel spin alignment - will not cause any demagnetization and has therefore not been observed in FM dynamics studies.

For discussing the FM case it is instructive to compare Dy with the neighboring lanthanide ferromagnet terbium (Tb), which has a very similar electronic structure. 
For Tb, two different channels transferring angular momentum from 4$f$ electrons to the lattice have been identified: i) via fast intra-atomic exchange with the delocalized 5$d$ valence electrons in the presence of hot electrons and ii) the slower direct 4$f$-spin-lattice coupling \cite{Wietstruk2011}. 
Interestingly, the fast decay channel i) is not found in our Dy FM data. 
Since structural and magnetic properties of Dy and Tb are very similar, major differences in 4$f$-5$d$ or 4$f$-lattice coupling are not to be expected. 
A main difference between the Tb experiment in Ref.\,\citenum{Wietstruk2011} and our Dy experiment is the sample thickness, though: the Tb film in Ref.\,\citenum{Wietstruk2011} was 10\,nm thick; while our Dy sample had a thickness of 120\,nm \cite{fn8}. 
It is to be expected that variation of the film thickness in this range (10\,nm are 35 monolayers) neither affects the 4$f$-5$d$ nor the 4$f$-lattice coupling.
On the other hand, spin transport should strongly depend on the sample dimensions as it involves spin currents into non-magnetic regions \cite{Battiato2010}. 
For our thick Dy film only the very thin non-magnetic cap layer is near the probed volume while the thick non-magnetic Y buffer layer is far away from the photoexcited regions. 
We therefore speculate that the fast time constant seen before in Tb may actually not be due to channel i) but rather be caused by spin transport. 

We would like to stress that the question about the existence of channel i) in FM Dy does not affect our conclusion about the interatomic spin transfer being fast and energy efficient: even if we missed a fast FM transfer channel in our Dy sample, this channel can be expected to have a similar time constant as the one in Tb. 
For the latter one, ($740\,\pm\,250$)\,fs was found \cite{Wietstruk2011,Eschenlohr2014}, which is still much slower than our result for the fast AFM dynamics in Dy.
We note that in a recent magneto-optical Kerr effect (MOKE) study of Dy a fast 300-fs~dynamics for the out-of-plane magnetization has been observed \cite{langner2015}. Similar time scales were also detected in MOKE experiments from FM Gd \cite{sultan2012} and were assigned to non-magnetic laser-induced changes of the optical sample properties \cite{koopmans2000,oppeneer2004} non-representative for the 4$f$-magnetic dynamics.

Coming back to the second, slower AFM dynamics with a time constant of ($22\,\pm\,7$)\,ps: this and the FM time constant of ($6\,\pm\,2$)\,ps are of similar order of magnitude as the time constant in Tb (8\,ps) related to the 4$f$-spin-lattice coupling [channel ii)] and should have the same origin. 
The quantitative difference we find between the two Dy phases hint to stronger 4$f$-lattice coupling in the FM phase \cite{Hubner1996}, which agrees with the observation that in FM Dy the 4$f$ spins are confined by a uniaxial in-plane anisotropy which is absent in the AFM phase \cite{jensen1991rare}.
We found no indications for an even slower AFM time scale of 200\,ps as reported by \citeauthor{langner2015} who studied the magnetic diffraction signal in Dy albeit with much lower temporal resolution of 70\,ps \cite{langner2015}.     

In Fig.\,\ref{fig:results}\,(d) we present an overview of the different angular momentum transfer channels with their characteristic time scales, including the interatomic spin transfer channel. 
In AFM Dy the interatomic transfer channel via hopping of 5$d$ electrons to adjacent atomic sites is effective in addition to those channels available in the FM phase. 
The opening of this channel leads to an up to 30 times faster reduction of the magnetic order compared to the FM phase.
For the 4$f$-5$d$ coupling we refer to the value of 10\,fs following Ref.\,\citenum{Ahuja1994} \cite{fs10}.

Our case study on Dy shows that for one and the same material the reduction of spin order is much faster and more energy-efficiently achieved when spins are antiferromagnetically aligned compared to FM spin order.
Generally any non-parallel spin alignment would allow to change the order parameter by redistributing angular momentum within the spin system. 
Since in the helical phase of Dy the angle between neighboring spins is only of the order of 34~degrees, even stronger effects may occur for larger relative angles. 
Our results apply primarily to $4f$ metals; since the angular momentum redistribution occurs through scattering of $5d$ electrons, similar effects can be expected as well in other systems where magnetic dynamics is dominated by $d$-electron scattering.

The highly efficient ultrafast interatomic transfer of angular momentum between non-parallel spins may define a route towards more energy-efficient ultrafast spin manipulation in devices. 
Non-parallel coupled magnetic moments may serve as spin sinks that reduce the energy required to manipulate spin order or allow for tuning time constants. 
The all-optical switching in, e.g., GdFeCo occurs via an almost complete quenching of the magnetization in the material \cite{Radu2011}. 
Most of the angular momentum needs to be transferred out of the 4$f$ system before switching sets in. 
Based on our finding, the energy needed to reach this transfer should be much lower when non-parallel 4$f$ spins are available either within the same material or possibly even in a multilayer structure.
Clever material design can make use of this effect to reduce the energy needed for ultrafast spin manipulation like optically induced magnetic switching.

N.T.-K., D.S., N.P., C.T., H.Z. and C.S.-L. planned and carried out the experiment with experimental support from R.M. and K.H.; H.Z. prepared the sample; N.T.-K., D.S. and N.P. analyzed the data; N.T.-K., N.P., D.S. and C.S.-L. wrote the manuscript with help from the other authors.

Funding: Work was supported by the BMBF (contract 05K10PK2). D.S. acknowledges the Helmholtz Association for funding via the Helmholtz Postdoc Program PD-142. N.T.-K. acknowledges the Helmholtz Association for funding via the Helmholtz Virtual Institute "Dynamic Pathways in Multidimensional Landscapes".


\begin{thebibliography}{38}
\bibitem[{fn8()}]{fn8}
	\bibinfo{note}{Thin Dy films do not develop a FM phase when
		epitaxially grown between Y layers.}
\end{thebibliography}

\end{document}


\title{Ultrafast and energy-efficient quenching of spin order:\\ Antiferromagnetism beats ferromagnetism}

\author{Nele Thielemann-K{\"u}hn}
\email{nele.thielemann@helmholtz-berlin.de}
\altaffiliation[Present address: ]{Fachbereich Physik, Freie Universit{\"a}t Berlin, Arnimallee 14, 14195 Berlin, Germany }%
\affiliation{Institut f{\"u}r Methoden und Instrumentierung der Forschung mit Synchrotronstrahlung, 
	Helmholtz-Zentrum Berlin f{\"u}r Materialien und Energie GmbH, Albert-Einstein-Stra{\ss}e 15, 12489 Berlin, Germany}%
\affiliation{Institut f{\"u}r Physik und Astronomie, Universit{\"a}t Potsdam, Karl-Liebknecht-Stra{\ss}e 24/25, 14476 Potsdam, Germany}%

\author{Daniel Schick}
\affiliation{Institut f{\"u}r Methoden und Instrumentierung der Forschung mit Synchrotronstrahlung, 
	Helmholtz-Zentrum Berlin f{\"u}r Materialien und Energie GmbH, Albert-Einstein-Stra{\ss}e 15, 12489 Berlin, Germany}%

\author{Niko Pontius}
\affiliation{Institut f{\"u}r Methoden und Instrumentierung der Forschung mit Synchrotronstrahlung, 
	Helmholtz-Zentrum Berlin f{\"u}r Materialien und Energie GmbH, Albert-Einstein-Stra{\ss}e 15, 12489 Berlin, Germany}%

\author{Christoph Trabant}
\affiliation{Institut f{\"u}r Methoden und Instrumentierung der Forschung mit Synchrotronstrahlung, 
	Helmholtz-Zentrum Berlin f{\"u}r Materialien und Energie GmbH, Albert-Einstein-Stra{\ss}e 15, 12489 Berlin, Germany}%
\affiliation{Institut f{\"u}r Physik und Astronomie, Universit{\"a}t Potsdam, Karl-Liebknecht-Stra{\ss}e 24/25, 14476 Potsdam, Germany}%
\affiliation{II. Physikalisches Institut, Universit{\"a}t zu K{\"o}ln, Z{\"u}lpicher Stra{\ss}e 77, 50937 K{\"o}ln, Germany}%

\author{Rolf Mitzner}
\affiliation{Institut f{\"u}r Methoden und Instrumentierung der Forschung mit Synchrotronstrahlung, 
	Helmholtz-Zentrum Berlin f{\"u}r Materialien und Energie GmbH, Albert-Einstein-Stra{\ss}e 15, 12489 Berlin, Germany}%

\author{Karsten Holldack}
\affiliation{Institut f{\"u}r Methoden und Instrumentierung der Forschung mit Synchrotronstrahlung, 
	Helmholtz-Zentrum Berlin f{\"u}r Materialien und Energie GmbH, Albert-Einstein-Stra{\ss}e 15, 12489 Berlin, Germany}%

\author{Hartmut Zabel}
\affiliation{Institut f{\"u}r Physik, Johannes-Gutenberg-Universit{\"a}t Mainz, Staudingerweg 7, 55128 Mainz, Germany}%

\author{Alexander F{\"o}hlisch}
\affiliation{Institut f{\"u}r Methoden und Instrumentierung der Forschung mit Synchrotronstrahlung, 
	Helmholtz-Zentrum Berlin f{\"u}r Materialien und Energie GmbH, Albert-Einstein-Stra{\ss}e 15, 12489 Berlin, Germany}%
\affiliation{Institut f{\"u}r Physik und Astronomie, Universit{\"a}t Potsdam, Karl-Liebknecht-Stra{\ss}e 24/25, 14476 Potsdam, Germany}%

\author{Christian Sch{\"u}{\ss}ler-Langeheine}
\affiliation{Institut f{\"u}r Methoden und Instrumentierung der Forschung mit Synchrotronstrahlung, 
	Helmholtz-Zentrum Berlin f{\"u}r Materialien und Energie GmbH, Albert-Einstein-Stra{\ss}e 15, 12489 Berlin, Germany}%
	
\date{\today}

\maketitle

\tableofcontents

\newpage

\section{Antiferromagnetic signal}

\begin{figure}[!htb]
    \makeatletter
    \renewcommand{\thefigure}{S\@arabic\c@figure}
    \makeatother

	\centering
	\includegraphics[width=0.9\textwidth]{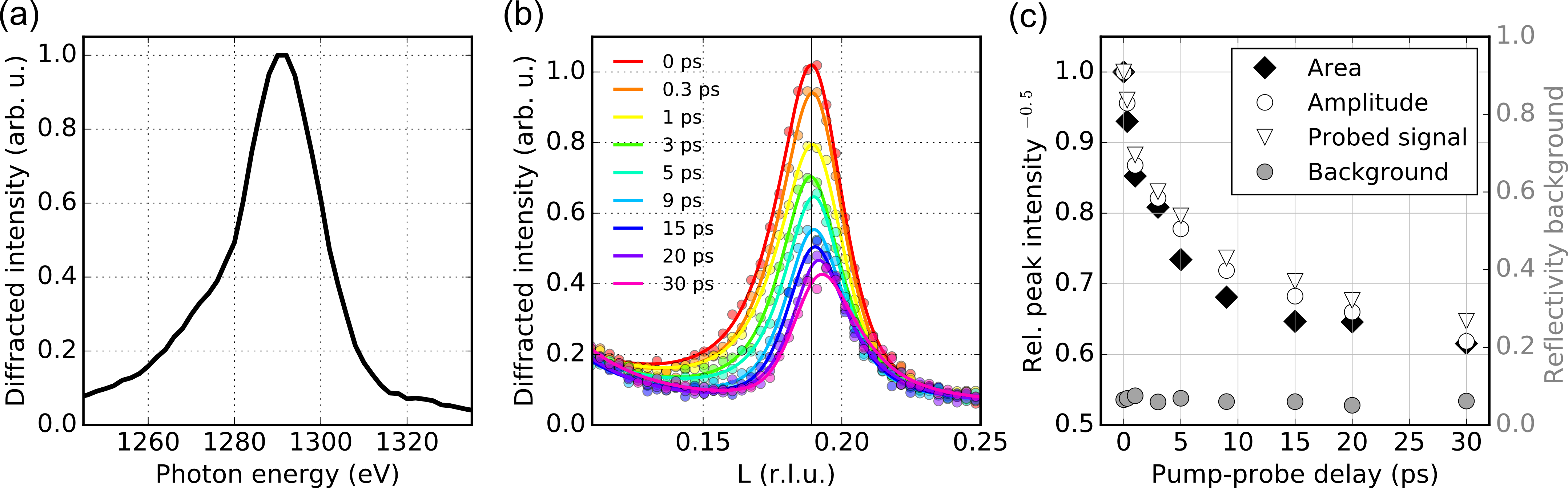}
	\caption{Dynamics of the magnetic Bragg peak.
		(a) Resonant x-ray diffraction spectrum around the Dy $M_5$-edge for momentum transfer kept at $(0\,0\,\tau)$.
		The spectrum was recorded at the FemtoSpeX slicing facility at BESSY\,II with an energy resolution of $E/\Delta E \approx 100$. 
		(b) Scans through momentum space along [0\,0\,1] ($L$) through the magnetic diffraction peak on the $M_5$-resonance for different pump-probe delays at an absorbed pump fluence of 1\,mJ/cm$^2$.
				The solid lines show the results of fits to the data. 
		The fits assume a pseudo-Voigt peak profile on a reflectivity background. 
		(c) Peak area and amplitude of the diffraction peak (minus reflectivity background), probed signal and reflectivity background at $\tau \approx 0.19$ for different pump-probe delays. 
		The quantities are determined from the fits to the diffraction peak in (b).
	}
	\label{fig:S1}
\end{figure}

The antiferromagnetic order parameter was probed via the intensity of the magnetic Bragg peak at (0\,0\,$\tau$). 
Figure\,\ref{fig:S1}\,(a) shows the x-ray resonant diffraction spectrum in the vicinity of the Dy $M_5$-resonance, i.e., the energy dependence of the magnetic peak. 
The intensity of this antiferromagnetic diffraction peak is proportional to the squared order parameter, i.e., $I\propto M^2$ \cite{Ott2006Ho}. 
At 110\,K the peak occurs at a momentum transfer of $\tau \approx0.19$ reciprocal lattice units (r.l.u.) along the [0\,0\,1] ($L$) direction. 

\noindent
The relation between $L$ and scattering angle $\theta$ is:
\begin{equation}
L = \frac{2c}{\lambda} \sin{\theta}
\end{equation}                                                                                                                                      
with $c = 5.654$\,{\AA} being the (room-temperature) lattice constant and $\lambda$ the photon wavelength. 
For 1292\,eV, $\lambda \approx 9.6$\,\AA.
For antiferromagnetic Dy, the peak position in momentum space depends sensitively on the sample temperature. We therefore verified that the peak position (and peak width) does not change within the first 15\,ps after photoexcitation [see Fig.\,\ref{fig:S1}\,(b)] such that the peak amplitude is proportional to the peak area and thus uniquely linked to the temporal evolution of the magnetic order.
Within the first 15\,ps we solely observe a reduction of the peak intensity.
For longer delays we also see a small shift of the peak to larger momenta. 
Peak shifts caused by static heating scale with the laser pump fluence and were taken into account by aligning the momentum transfer, accordingly.
Besides the antiferromagnetic signal, the scattered intensity probed at the peak maximum position contains a background signal caused by specular reflectivity. 
This background is not expected to be affected by pumping. 
Since we did not determine the reflectivity background for every delay scan we underestimate the pump effect for the antiferromagnetic case slightly. 
This is illustrated in Fig.\,\ref{fig:S1}\,(c).

\newpage

\section{Ferromagnetic signal}

\begin{figure}[!h]
    \makeatletter
    \renewcommand{\thefigure}{S\@arabic\c@figure}
    \makeatother
	\centering
	\includegraphics[width=0.7\textwidth]{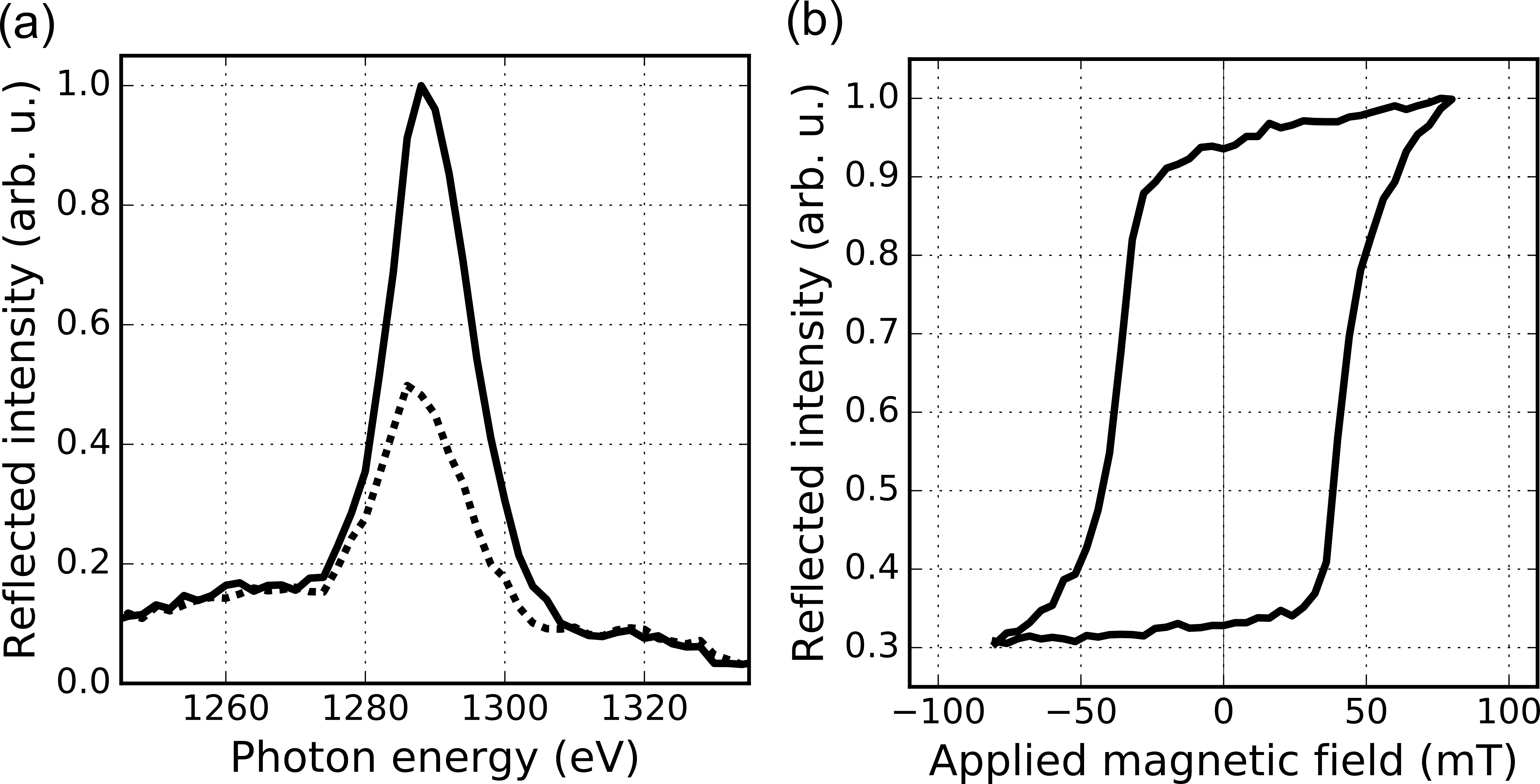}
	\caption{Magnetic circular dichroism (MCD) in reflection. 
		(a) X-ray reflection spectrum of the Dy $M_5$-edge in the ferromagnetic phase recorded with circularly polarized x-rays for opposite magnetization directions (solid and dashed line). 
		(b) Magnetic hysteresis measured at a fixed photon energy of 1286\,eV.
	}
	\label{fig:S2}
\end{figure}

In the ferromagnetic phase we recorded the difference in specular reflected intensity at the Dy $M_5$-resonance for different relative orientations of magnetization and x-ray helicity. This quantity is proportional to the magnetization \cite{Mertins2002}. 
The magnitude of the magnetic circular dichroism in reflection geometry depends on the incidence angle. 
We chose an incidence angle of 5\,degrees with respect to the sample surface, at which we reach large magnetic contrast with high overall intensity [Fig.\,\ref{fig:S2}\,(a)]. 
To magnetize the sample a static magnetic field of 80\,mT was applied alternating between essentially parallel and antiparallel orientation with respect to the fixed x-ray helicity vector (in-plane magnetization). 
The magnetic hysteresis measured with this field geometry is shown in the Fig.\,\ref{fig:S2}\,(b). 

\newpage

\section{Long exponential time constants}

\begin{figure}[!h]
    \makeatletter
    \renewcommand{\thefigure}{S\@arabic\c@figure}
    \makeatother
	\centering
	\includegraphics[width=0.35\textwidth]{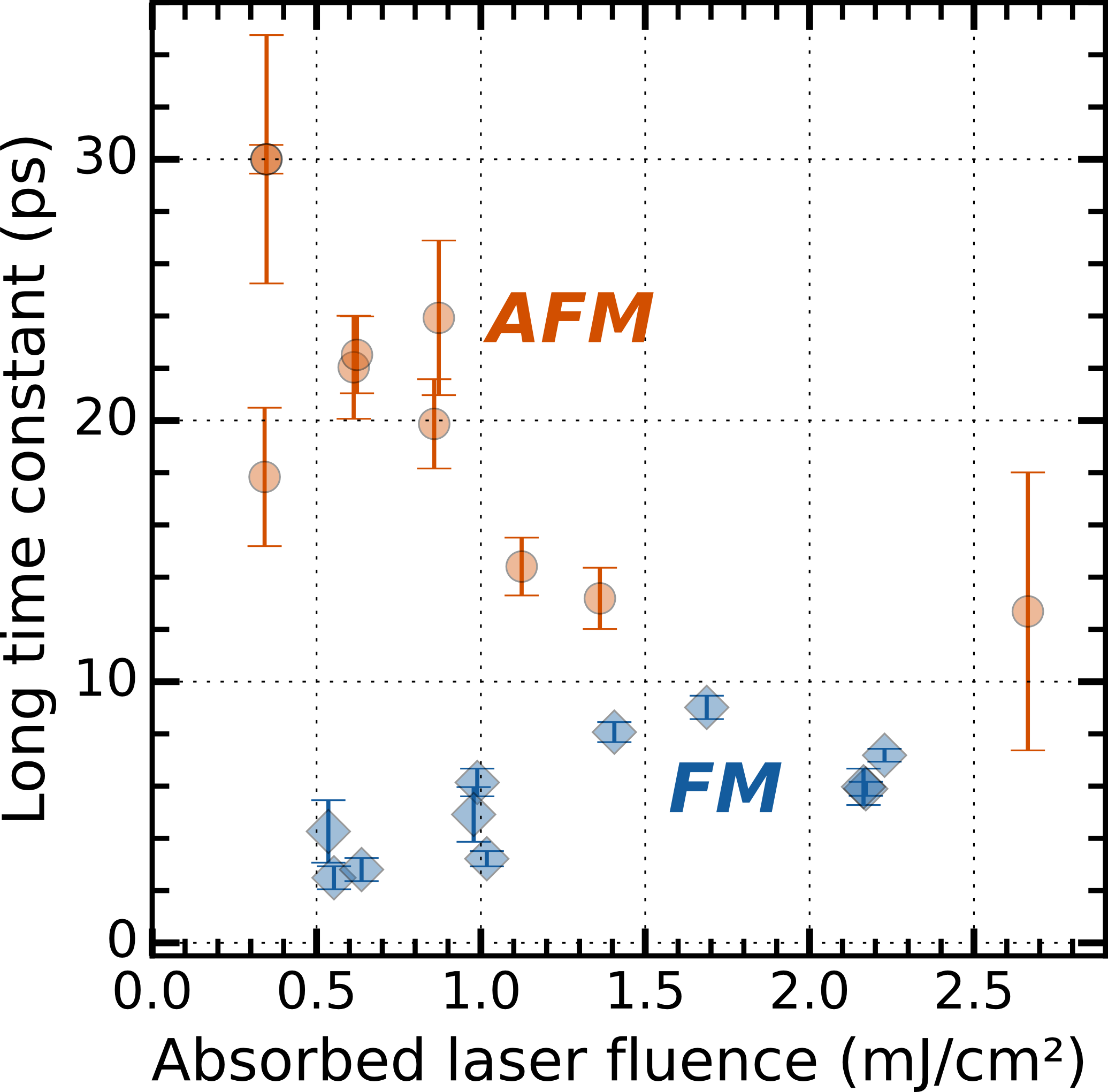}
	\caption{Long exponential time constants extracted via the least-square fits as shown in Fig.\,S5\&S6. 
		We determined an average slow exponential time constant of ($22\,\pm\,7$)\,ps for the antiferromagnetic (orange circles) and a single time constant of ($6\,\pm\,2$)\,ps for the ferromagnetic phase (blue diamonds). 
	}
	\label{fig:S3}
\end{figure}

\section{Data acquisition and treatment}

\begin{figure}[!htb]
    \makeatletter
    \renewcommand{\thefigure}{S\@arabic\c@figure}
    \makeatother
	\centering
	\includegraphics[width=0.6\textwidth]{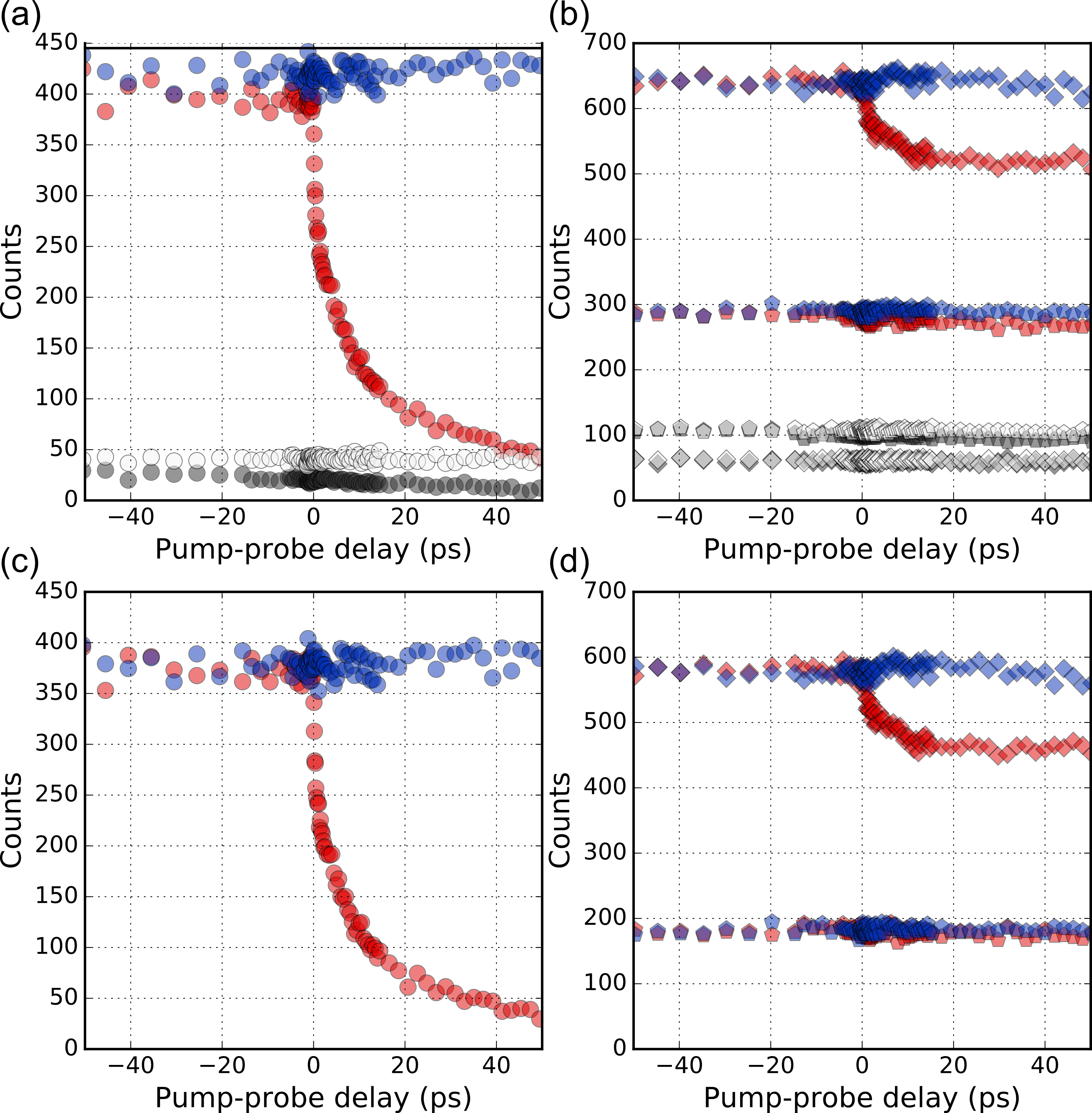}
	\caption{Raw data and background subtraction. 
		(a) \& (b) Summed up raw data for a set of delay scans for a particular pump fluence (1\,mJ/cm$^2$). 
		The \emph{pumped} signal is denoted with red and the \emph{unpumped} signal with blue symbols.
		The respective halo backgrounds are the dark grey symbols (pumped) and light grey symbols (unpumped). 
		Plot (a) exhibits the raw data for the antiferromagnetic case and (b) for the ferromagnetic case. 
		In the latter the diamonds denote positive and the pentagons negative magnetization direction.
		(c) \& (d) Respective delay scans after background subtraction. 
	}
	\label{fig:S4}
\end{figure}

The x-ray photons from the sample were detected with an avalanche photodiode (APD, Laser-Component SAR3000) operated close to the breakthrough voltage.
This allowed us to use single photon counting detection. 
The APD is screened from the laser light by a 250\,nm thick aluminum membrane which is attached to the light tight aluminum housing of the APD.
Besides the dynamic magnetic signal with $\approx 120$\,fs temporal resolution, the raw detector signal contains background contributions from the halo background of the x-ray slicing source. 
The halo background is a consequence of the repetitive excitation of an identical electron bunch from the storage ring by the slicing process \cite{Holldack2014}. 
This results in an x-ray radiation background pulse with 70\,ps duration. 
For the present experiment the ratio between halo and femtosecond x-ray intensity was typically 1/10 for linear polarization and 1/5 for elliptical polarization.
To reliably eliminate the halo background from the data we directly measured separately the delay dependent transient of the halo background without the fs-x-ray pulses \cite{Schick2016} for the pumped and the unpumped sample. 
The halo contribution (which also contains possible other background contributions due to dark counts in the detector or due to electronic noise) is then subtracted from the transient signal recorded with the femtosecond x-ray pulses yielding the background-free \emph{pumped} and \emph{unpumped} signals that we used for the quantitative analysis (Fig.\,\ref{fig:S4}).

\newpage

\section{Data analysis}

For both magnetic phases we recorded the \emph{pumped} delay traces, the corresponding \emph{unpumped} signal and the respective halo background. 
Figures\,\ref{fig:S4}\,(a)\&(b) show raw data for 1\,mJ/cm$^2$ absorbed laser fluence. 
After background subtraction [Fig.\,\ref{fig:S4}\,(c)\&(d)] the signal proportional to the amount of magnetic order for a given delay is calculated for the antiferromagnetic case as
\begin{equation}
S_\text{AFM} = \sqrt{\frac{S^p}{S^u}}
\end{equation}
where $S^p$ and $S^u$ are background corrected \emph{pumped} and \emph{unpumped} signal, respectively. 
For the measurements in the ferromagnetic case the signal providing the amount of magnetic order is calculated as
\begin{equation}
S_\text{FM} = \frac{\left( \overline{S^u} + S^p - S^u \right)^-  - \left( \overline{S^u} + S^p - S^u  \right)^+}{\overline{S^u}^- - \overline{S^u}^+}
\end{equation}
Here, $\overline{S^u}$ is the delay-scan averaged \emph{unpumped} signal.
The antiferromagnetic transient signal $F_\text{AFM}(t)$ is fitted by a double exponential decay function yielding two exponential time constants ($f$=fast, $s$=slow) convoluted by a Gaussian function to account for the temporal resolution:
\begin{equation}
F_\text{AFM}(t) = f_\text{AFM}(t) \otimes \text{Gauss}(\Delta t)
\end{equation}
with
\begin{equation}
\begin{array}{r @{} c @{} l @{} }
f_\text{AFM}(t) &{}=\displaystyle
\begin{cases}
I_0 & \text{, } t \leq t_0,\\
I_0 - I_f \left( 1- e^{-\frac{t-t_0}{\tau_f}} \right) 
- I_s \left( 1- e^{-\frac{t-t_0}{\tau_s}} \right) & \text{, } t > t_0.
\end{cases}
\end{array}
\end{equation}
The ferromagnetic dynamics were fitted by a single exponential decay function,
\begin{equation}
\begin{array}{r @{} c @{} l @{} }
f_\text{FM}(t) &{}=\displaystyle
\begin{cases}
I_0 & \text{, } t \leq t_0,\\
I_0 - I_s \left( 1- e^{-\frac{t-t_0}{\tau_s}} \right) & \text{, } t > t_0.
\end{cases}
\end{array}
\end{equation}
again convoluted by a Gaussian function.

\begin{figure*}
    \makeatletter
    \renewcommand{\thefigure}{S\@arabic\c@figure}
    \makeatother
	\centering
	\includegraphics[width=0.9\textwidth]{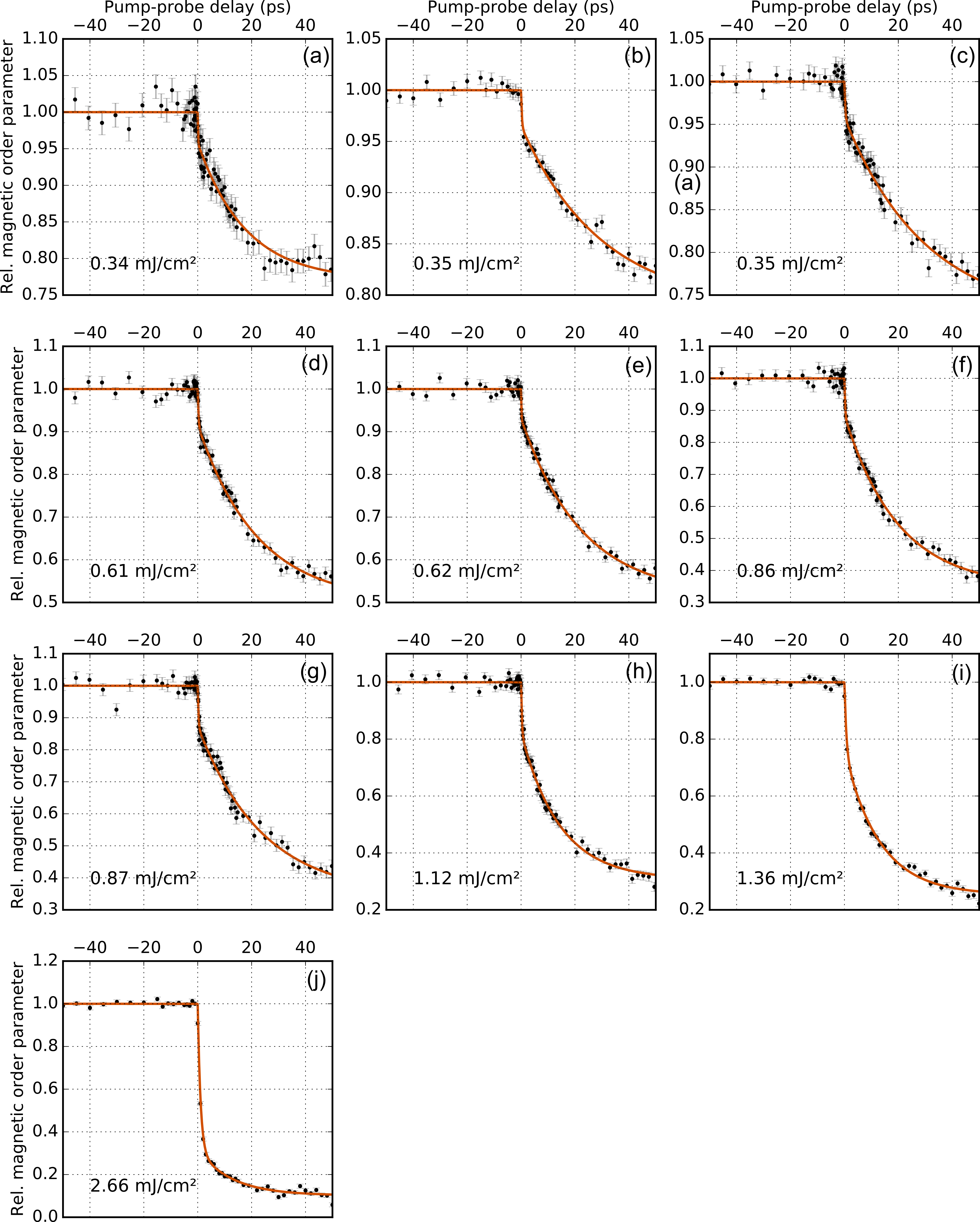}
	\caption{Pump-probe delay scans for antiferromagnetic order. 
		For the relative order parameter the average signal at negative delay is normalized to 1. 
		The error bars correspond to the one-$\sigma$ standard deviation determined at negative delay.
		The solid lines denote the results of the fit analysis as described above.}
	\label{fig:S5}
\end{figure*}


\begin{figure*}
    \makeatletter
    \renewcommand{\thefigure}{S\@arabic\c@figure}
    \makeatother
	\centering
	\includegraphics[width=0.9\textwidth]{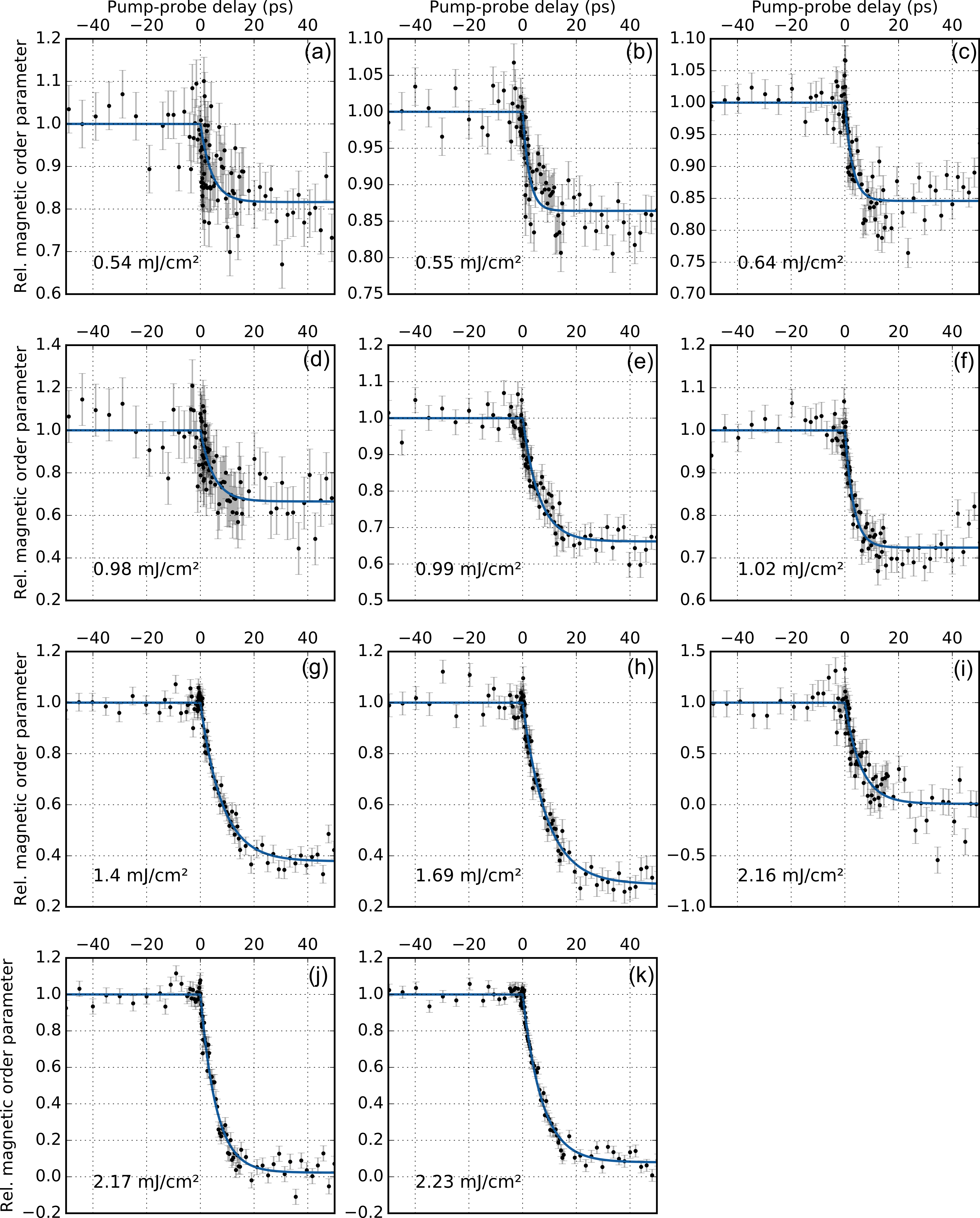}
	\caption{Pump-probe delay scans for ferromagnetic order. 
		For the relative order parameter the average signal at negative delay is normalized to 1. 
		The error bars correspond to the one-$\sigma$ standard deviation determined at negative delay.
		The solid lines denote the results of the fit analysis as described above.  
	}
	\label{fig:S6}
\end{figure*}

\clearpage

\section{Statistical Analysis}

All pump-probe delay scans for either the ferromagnetic or antiferromagnetic signals at a given pump fluence were normalized such that the average signal at negative delay was 1. 
The error bars in Fig.\,\ref{fig:S5}\&\ref{fig:S6} correspond to the one-$\sigma$ standard deviation of the data points at this negative delay. 
We verified that the such determined error bars were of the same size as when determined from counting statistics. 
All other quantities including parameter errors were derived from least-squares fits to the data. 
The fitting was performed with the lmfit package \cite{Newville2014}. 
After having verified, that $t_0$ did not change between the experiments in the ferromagnetic and antiferromagnetic phase, we analyzed the data with $t_0$ fixed to a value determined from pump-probe delay scans performed at high laser fluences, where this quantity can be determined most accurately. 

\section{Angular momentum transfer rate}

As momentary angular momentum transfer rate we define the change of magnetic order parameter per time, hence the rate of transfer of angular momentum out of the ordered $4f$ spin system. 
We determined the angular momentum transfer rate from the fits to the delay scans by deriving them and scaling them to the magnitude of the equilibrium order parameter $M(T)$ at the particular sample temperature $T$. 
From Ref.\,\citenum{Wilkinson1961} we inferred an ordered $4f$ magnetic moment of 9.84\,$\upmu_B$ at 40\,K and 8.61\,$\upmu_B$ at 110\,K. 
The maximum angular momentum transfer rate is the maximum change of magnetic order parameter for all pump-probe delays. 

\newpage

\section{Pump fluence determination}

\begin{figure}[!htb]
    \makeatletter
    \renewcommand{\thefigure}{S\@arabic\c@figure}
    \makeatother
	\centering
	\includegraphics[width=0.4\textwidth]{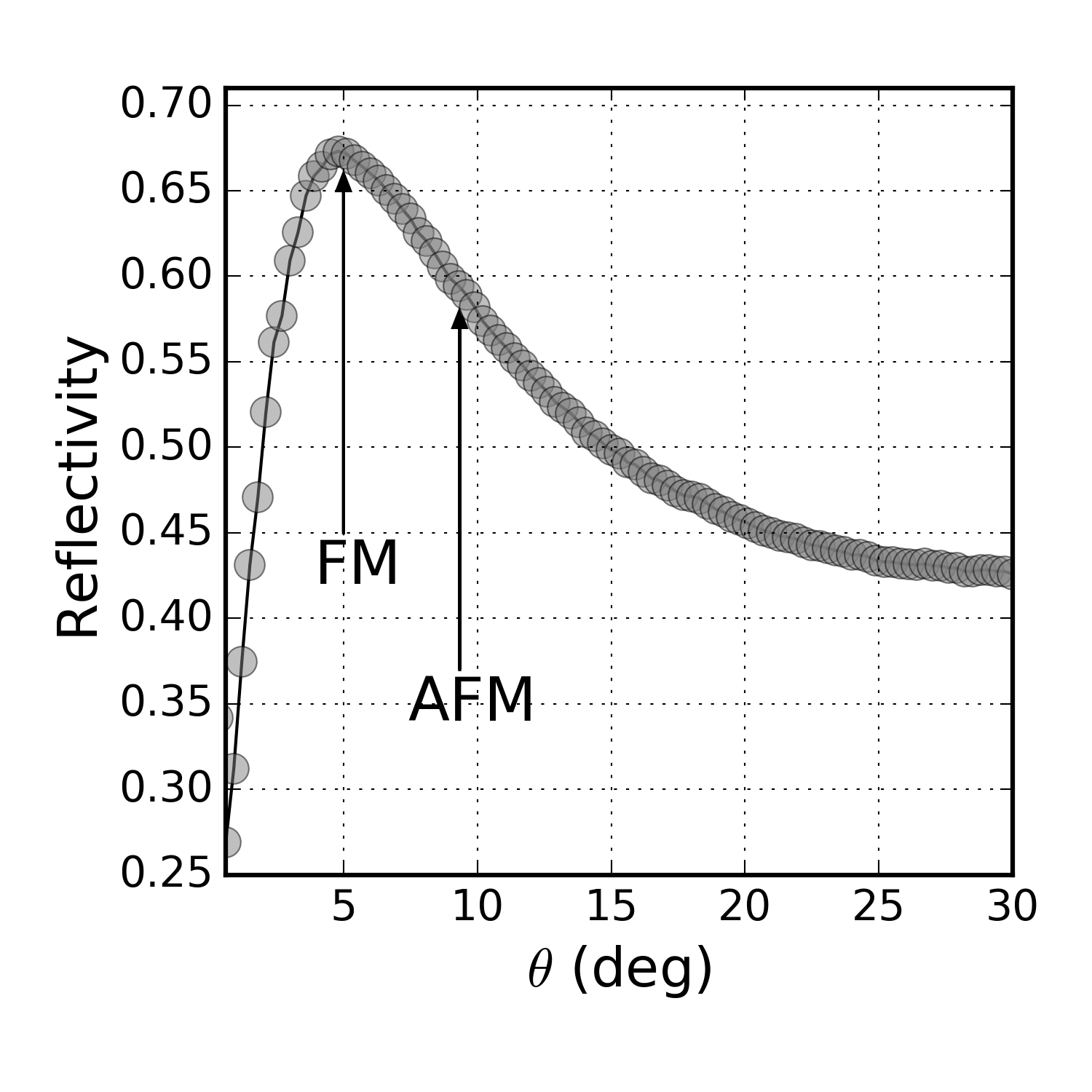}
	\caption{Surface reflectivity from the Dy sample of the 800\,nm pump-laser beam as a function of incidence angle $\theta$ with respect to the sample surface. 
	The arrows indicate the angles at which the time-resolved x-ray measurements were carried out for the ferromagnetic and antiferromagnetic case, respectively. 
	The corresponding pump-laser-reflectivity values were used for determining the absorbed laser fluence. 
	The drop off of reflected intensity for shallow angles below $\approx$ 7\,degrees is a consequence of the projected laser beam spot which is getting larger than lateral dimensions of the sample leading to part of the laser beam missing the sample. 
	}
	\label{fig:S7}
\end{figure}

The pump-laser fluence given in the present study is the absorbed fluence taking into account footprint effect and surface reflectivity. 
The pump-laser-pulse energy is determined from the laser power immediately before the laser is coupled into the vacuum chamber of the experiment. 
The laser-spot size on the sample is experimentally determined in situ by scanning a 50\,$\upmu$m pinhole through the laser beam at the sample position. 
The pump-laser reflectivity from the sample surface was determined experimentally as well (see Fig.\,\ref{fig:S7}). 

\noindent
The footprint effect due to the grazing incidence enters by the factor $1/\sin{\theta}$. 
The absorbed laser fluence is then calculated as
\begin{equation}
F_l = \frac{P}{A\cdot R} \sin{\theta} (1-r)
\end{equation}
where $P$ is the laser power, $R$ the repetition rate, $A$ the laser cross section at the sample position, $\theta$ the scattering angle and $r$ the relative reflectivity (Fig.\,\ref{fig:S7}).
For measurements with varying scattering angle $\theta$, the laser power $P$ was adapted for every data point to compensate a change of the factor $\sin{\theta} (1-r)$ such that the absorbed fluence $F_l$ was kept at a constant value.

\section{Optical-pump and x-ray-probe penetration depth}

In order to assure that the probed volume in the sample is fully excited by the pump laser we determined the penetration depth for the optical pump pulse and the x-ray-probe pulses considering refraction. From the absorption coefficient and refraction index for 1.5\,eV photons in single crystalline Dy ($E\parallel c$) given in Ref.\,\citenum{Adachi2012}, it follows an optical penetration depth of $l_{IR}\approx 21\,\mbox{nm}$.
The probing depth for resonant x-rays with energies tuned to the M$_5$-absorption edge is much shorter \cite{Ott2010}. For our experimental condition (scattering geometry, x-ray band width) we find an x-ray penetration depth for the ferromagnetic case of ${l_{x}(\theta=5)\approx 7\,\mbox{nm}}$ and ${l_{x}(\theta=9.5)\approx 12\,\mbox{nm}}$ for the antiferromagnetic case.

\vspace{1cm}

All plots in the Manuscript and the Supplementary Materials were created with the matplotlib Python package \cite{Hunter}.

\newpage

